%% file: MU-MIMO.tex
\documentclass[12pt, draftclsnofoot, onecolumn, comsoc]{IEEEtran}

\usepackage{amsmath}
\usepackage{amssymb}
\usepackage{mathtools}
\usepackage{mathdots}
\usepackage[caption=false]{subfig}
\usepackage{graphicx}
\usepackage{epstopdf}
\usepackage{cite}
\usepackage{relsize}
\usepackage{mathrsfs}

\DeclareMathOperator{\Tr}{Tr}
\DeclareMathOperator{\diag}{diag}

\DeclareMathOperator{\nulls}{null}

\DeclareMathOperator{\rank}{rank}

\DeclareMathOperator{\sinr}{SINR}

\DeclareMathOperator{\eq}{EQ}
\DeclareMathOperator{\jp}{JP}
\DeclareMathOperator{\timer}{TR}

\DeclareMathOperator{\dd}{D}
\DeclareMathOperator{\isi}{ISI}
\DeclareMathOperator{\iui}{IUI}
\DeclareMathOperator{\nn}{N}
\DeclareMathOperator{\qam}{QAM}

\newcommand{\E}{\mathbb{E}}
\newcommand{\C}{\mathbb{C}}
\newcommand{\R}{\mathbb{R}}

\newcommand{\lag}{\mathcal{L}}

\newcommand{\HH}{\mathbf{H}}
\newcommand{\PP}{\mathbf{P}}

\newcommand{\V}{\mathbf{V}}
\newcommand{\A}{\mathbf{A}}
\newcommand{\B}{\mathbf{B}}
\newcommand{\CC}{\mathbf{C}}

\newcommand{\G}{\mathbf{G}}
\newcommand{\I}{\mathbf{I}}
\newcommand{\Q}{\mathbf{Q}}

\newcommand{\U}{\mathbf{U}}

\newcommand{\aaa}{\mathbf{a}}
\newcommand{\h}{\mathbf{h}}

\newcommand{\s}{\mathbf{s}}

\newcommand{\y}{\mathbf{y}}
\newcommand{\z}{\mathbf{z}}

\newcommand{\Gb}{\bar{\mathbf{G}}}

\newcommand{\Htil}{\tilde{\mathbf{H}}}
\newcommand{\Util}{\tilde{\mathbf{U}}}
\newcommand{\Vtil}{\tilde{\mathbf{V}}}
\newcommand{\Sigtil}{\tilde{\boldsymbol{\Sigma}}}

\newcommand{\gu}{\breve{g}}

\newcommand{\Gu}{\mathbf{\breve{G}}}
\newcommand{\Pu}{\mathbf{\breve{P}}}

\newcommand{\Hbar}{\bar{\mathbf{H}}}
\newcommand{\Pb}{\mathbf{\bar{P}}}

\newcommand{\rhob}{\boldsymbol{\rho}}

\newcommand{\Lam}{\boldsymbol{\Lambda}}
\newcommand{\Sig}{\boldsymbol{\Sigma}}

\newcommand{\ze}{\mathbf{0}}

\begin{document}
\title{Space-Time Block Diagonalization for Frequency-Selective MIMO Broadcast Channels}
\author{Carlos A. Viteri-Mera and Fernando L. Teixeira
\thanks{This work was supported by a Fulbright Colombia fellowship. Parts of this work were submitted to the IEEE Global Communications Conference 2016 \cite{viteri2016}.}%
\thanks{The authors are with the ElectroScience Laboratory, Department of Electrical and Computer Engineering, The Ohio State University, Columbus, OH (email: \{viteri.5; teixeira.5\}@osu.edu).}%
\thanks{C. Viteri-Mera is also with the Department of Electronics Engineering, Universidad de Nari\~no, Pasto, Colombia.}%
}%
\maketitle

\vspace*{-24pt}
\begin{abstract}
The most relevant linear precoding method for frequency-flat MIMO broadcast channels is block diagonalization (BD) which, under certain conditions, attains the same nonlinear dirty paper coding channel capacity. However, BD is not easily translated to frequency-selective channels, since space-time information is required for transceiver design. In this paper, we demonstrate that BD is feasible in frequency-selective MIMO broadcast channels to eliminate inter-user interference (IUI) if the transmit block length is sufficiently large, and if the number of transmit antennas is greater than the number of users. We also propose three different approaches to mitigate/eliminate inter-symbol interference (ISI) in block transmissions: \emph{i}) time-reversal-based BD (TRBD) which maximizes spatial focusing around the receivers using transmitter processing only, \emph{ii}) equalized BD (EBD) which minimizes the ISI using transmitter processing only, and \emph{iii}) joint processing BD (JPBD), which uses linear processing at the transmitter and the receiver to suppress ISI. We analyze the theoretical diversity and multiplexing gains of these techniques, and we demonstrate that JPBD approximates full multiplexing gain for a sufficiently large transmit block length. Extensive numerical simulations show that the achievable rate and probability of error performance of all the proposed methods improve those of conventional time-reversal beamforming. Moreover, JPBD provides the highest achievable rate region for frequency-selective MIMO broadcast channels.
\end{abstract}

\input{Introduction}

\input{SystemModel}

\input{BDPrecoding}

\input{JointProcessing}

\input{PowerAllocation}

\input{Analysis}

\input{Results}

\input{Conclusions}

\input{Appendices}


\bibliographystyle{IEEEtran}
\bibliography{MU-MIMO}

\end{document}

%% file: Introduction.tex
\section{Introduction}
Wireless multiuser-MIMO (MU-MIMO) systems are composed of a multiple-antenna base station and a set of user terminals (possibly, but not necessarily, equipped with multiple antennas). In the downlink, the system is modeled as a MIMO broadcast channel where each user receives a linear combination of the signals directed to all the users. Thus, the main characteristic of these systems is the presence of inter-user interference (IUI) and, as a result, processing techniques at the transmitter and/or receiver are required so that every user can detect the signal directed to it. A number of such methods exist which operate on different principles depending on the channel being frequency-flat or frequency selective. Dirty paper coding (DPC), a nonlinear method, achieves the capacity in frequency-flat MIMO broadcast channels \cite{weingarten2006,hassibi2007}. For frequency-selective channels, the capacity region is unknown in terms of the channel statistics, even in the SISO scenario \cite{tulino2010}.

Despite the fact that DPC achieves capacity in frequency-flat MIMO broadcast channels, linear processing techniques are of great interest since they offer reduced computational complexity compared to DPC \cite{spencer2004,peel2005,gesbert2007}. In particular, block diagonalization (BD) \cite{spencer2004} is of significant interest given that, under certain conditions, it achieves the DPC sum capacity~\cite{shen2007}. BD uses a linear precoder to set the IUI to zero, which forces a block-diagonal structure in the precoder-channel matrix product. In frequency-flat channels, the channel matrix has only space information (the complex channel coefficients between each transmitter/receiver antenna pair). For frequency-selective channels, the channel matrix incorporates space-time information since a channel impulse response (CIR) characterizes the propagation between each transmitter/receiver antenna pair. Hence, frequency-flat linear processing techniques are not easily extended to the frequency-selective case.

The main challenge in frequency-selective channels is the presence of inter-symbol interference in the received signal caused by time-domain spread. Thus, the transmitter and/or the receivers must use equalization in order to mitigate ISI. For the specific case of frequency-selective MIMO broadcast channels, time-reversal (TR) based pre-filters~\cite{lerosey2005,yavuz2008,fouda2012,han2012} have been extensively used since they improve the system's energy efficiency and reduce its computational complexity with respect to multicarrier (frequency-flat) systems~\cite{chen2013}. TR uses the time-reversed complex-conjugated CIR as a linear pre-filter applied at the transmitter, and uses simple single-tap receivers. TR focuses the electric field around the receiving antennas \cite{yavuz2005} and also provides partial equalization due to its matched-filter properties, compressing the equivalent CIR in the time-domain~\cite{viteri2014}. However, TR performance is limited by both ISI and IUI \cite{viteri2015}, so the design of linear processing techniques in frequency-selective MIMO broadcast channels is still an open problem.

In this work, we generalize BD linear precoding to frequency-selective MIMO broadcast channels. We show that BD is possible in this case if the transmitted block length is sufficiently large and if the number of transmit antennas is greater than the number of users (or equal to, in some cases). The processing in frequency-selective channels involves space-time information, and we show that any BD precoder in this case acts as a space-time block coder that eliminates IUI. In addition, we propose three approaches to mitigate or eliminate ISI in the received signal, which work in cascade configuration with the BD precoder. The first two approaches, time-reversal-based BD (TRBD) and equalized BD (EBD) use channel state information (CSI) at the transmitter only to design linear precoders and use low complexity sample-drop receivers. The third approach, joint processing BD (JPBD) uses CSI at the transmitter and the receivers to jointly calculate linear precoders and receiver combiners.

TRBD is based on the frequency domain formulation proposed in \cite{viteri2015}, where IUI is eliminated (with a BD precoder) and ISI is mitigated by approximating the TR pre-filter. The second approach, EBD, acts explicitly as a pre-equalizer \cite{proakis2008} over each block-diagonalized channel, giving a minimum squared error solution for the precoder. JPBD uses the singular value decomposition (SVD) of the block-diagonalized channel to eliminate ISI and provides perfect equalization in the received signal. For each approach, we theoretically analyze:
\begin{enumerate}
\item The optimization problems related to the precoder design, which have closed-form solutions in each case.
\item The ergodic achievable rate region.
\item The high SNR performance, evaluated in terms of the diversity and multiplexing gains.
\item The effective signal to interference plus noise ratio (SINR) for low SNR.
\end{enumerate}
Extensive numerical numerical simulations show that the achievable rate regions of the proposed techniques improve those of conventional TR beamforming. Moreover, we demonstrate that any linear precoding technique (processing at the transmitter only, including TRBD and EBD) cannot eliminate ISI completely, implying zero diversity and multiplexing gains. JPBD achieves full multiplexing gain (equal to the number of users) in the limit when the transmitted block size goes to infinity, and its diversity gain improves with larger channel delay spreads or larger time-domain redundancy added at the transmitter. With these characteristics, JPBD provides the highest known achievable rate region for frequency-selective MIMO broadcast channels. We also analyze the behavior of each design versus different system parameters (e.g. number of antennas, number of users, SNRs) and show good agreement between simulated and theoretical results.

\begin{figure*}[t]
\centering
\includegraphics[width=\columnwidth]{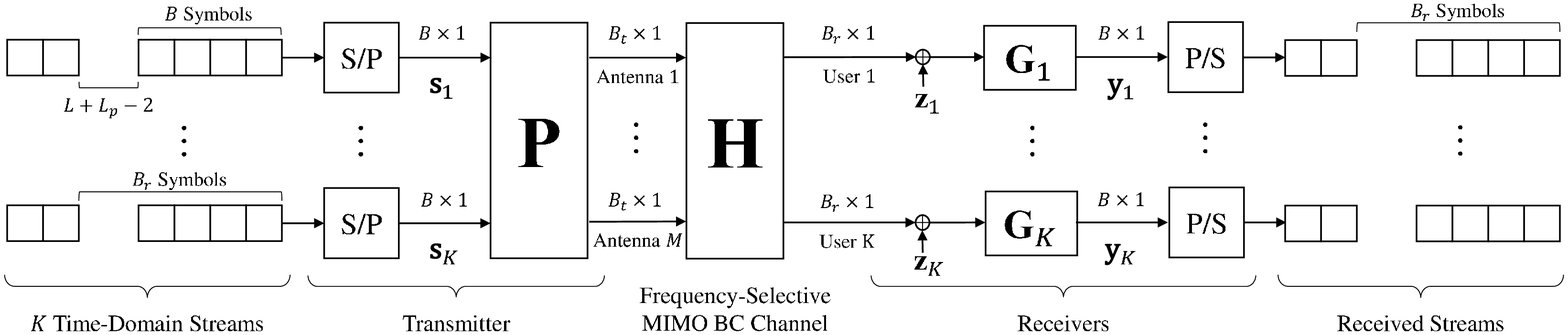}
\caption{Frequency-selective MU-MIMO downlink model.}
\label{fig_system}
\end{figure*}

%% file: SystemModel.tex
\section{System Model}
Consider a MIMO baseband \emph{downlink} wireless communication system consisting of one transmitter (base station or access point) equipped with $M$ transmit antennas and $K$ single-antenna users, as depicted in Fig. \ref{fig_system}. The system operates over a MU-MIMO fading channel, where the transmitter sends a block of $B$ complex symbols to each user, followed by a guard interval of  $L+L_p-2$ symbols, where $L$ is the delay spread in the channel and $L_p$ is the redundancy added by the precoder. At user $k$, the received signal is represented as
\begin{IEEEeqnarray}{rCl}
\label{eq_received}
\y_k = \G_k \left[ \HH_{k} \PP_{k} \s_k + \sum_{k'=1 , k' \neq k}^K  \HH_{k} \PP_{k'} \s_{k'} + \z_k \right],
\end{IEEEeqnarray}
where $\G_k$ is the receiver filter, $\HH_{k}$ is the channel matrix, $\PP_{k}$ is the transmitter precoder, $\s_k$ is the transmitted signal, and $\z_k$ is Gaussian noise. In this section, we describe this system model in detail\footnote{We use the following notation. $(\cdot)^T$, $(\cdot)^*$, $(\cdot)^H$, $(\cdot)^{-1}$, $(\cdot)^+$, and $\Vert \cdot \Vert_F$ represent transpose, complex conjugate, conjugate transpose, inverse, pseudoinverse, and Frobenius norm of a matrix, respectively. $[\A]_{ij}$ is the element in the $i$-th row and $j$-th column of matrix $\A$. $\Vert \aaa \Vert_2 = \sqrt{\aaa^H\aaa}$ is the $\ell_2$ norm of the vector $\aaa$. $\E\{\cdot\}$ denotes expected value. We use the definitions in \cite{hjorungnes2007} for complex matrix differentiation.}.

\subsection{Transmitter}
Let $\s_k = [s_k(1), \ldots , s_k(B)]^T \in \C^B$ denote the random vector of complex time-domain transmitted symbols, where $s_k(t)$ is the symbol directed to user $k$ at time $t$ with average power $\E \left\{ \vert s_k(t) \vert^2 \right\} = \rho_k$, $\forall t$. These time domain symbols are i.i.d. random variables selected from an arbitrary alphabet. As shown in Fig. \ref{fig_system}, the precoding matrix $\PP$ maps the stacked transmitted signal vector $\s = \left[ \s_1^T,\ldots,\s_K^T \right]^T \in \C^{BK}$ to the transmit antennas. The total transmitted power constraint is $\sum_{k} \rho_k = P_{\max}$, and the precoding matrix $\PP \in \C^{M(B+Lp-1) \times BK}$ is
\begin{IEEEeqnarray}{rCl}
\PP & = &
\begin{bmatrix}
\PP_{1,1}	&	\PP_{1,2}	&	\cdots	&	\PP_{1,K}	\\
\PP_{2,1} 	&	\PP_{2,2}	& 	\cdots	&	\PP_{2,K}	\\
\vdots		&	\vdots		&	\ddots	&	\vdots		\\
\PP_{M,1} 	&	\PP_{M,2} 	&	\cdots	&	\PP_{M,K}	\\
\end{bmatrix} \nonumber
\end{IEEEeqnarray}
where $\PP_{m,k} \in \C^{(B+L_p-1) \times B}$ is the linear combiner which maps the time-domain block ($B$ symbols) directed to user $k$ to a time-domain block transmitted from antenna $m$ ($B_t=B+L_p-1$ symbols). Thus, the precoders add $L_p-1$ time-domain redundancy symbols. Note that, when the precoder is a finite impulse response (FIR) filter of length $L_p$, $\PP_{m,k}$ is a banded Toeplitz matrix representing the convolution between the filter and the transmitted block \cite{hayes1996}. We define $\PP_k = \left[ \PP_{1,k}^T \cdots \PP_{M,k}^T \right]^T \in \C^{B_t M \times B}$ as the stacking of all the precoders directed to user $k$, such that $\PP = \left[ \PP_1 \cdots \PP_K \right]$. We also set $\Vert \PP_k \Vert_F^2 = 1$, $\forall k$, so the combiner does not alter the average power of $\s_k$. Given the previous definitions, $\PP$ is a linear space-time block coder.

\subsection{Channel}
\label{sec_channel}
We focus on quasi-static channels, where the channel matrix remains invariant over a block of $B+L+L_p-2$ time samples. The frequency-selective MIMO broadcast channel matrix $\HH \in \C^{K(B_t + L - 1) \times B_t M}$ is
\begin{IEEEeqnarray}{rCl}
\HH = 
\begin{bmatrix}
\HH_{1,1}	&	\HH_{1,2}	&	\cdots	&	\HH_{1,M}	\\
\HH_{2,1} 	&	\HH_{2,2}	& 	\cdots	&	\HH_{2,M}	\\
\vdots		&	\vdots		&	\ddots	&	\vdots		\\
\HH_{K,1} 	&	\HH_{K,2} 	&	\cdots	&	\HH_{K,M}	\\
\end{bmatrix} =
\begin{bmatrix}
\HH_{1}	\\
\HH_{2} 	\\
\vdots		\\
\HH_{K} 	\\
\end{bmatrix}, \nonumber
\end{IEEEeqnarray}
where $\HH_{k,m} \in \C^{(B_t+L-1) \times B_t}$ is a banded Toeplitz convolution matrix with the CIR coefficients from transmit antenna $m$ to user $k$ given by
\begin{IEEEeqnarray}{rCl}
\HH_{k,m} = 
\begin{bmatrix}
h_{k,m}(1)	&	0				&	\cdots	&	0			\\
\vdots		&	h_{k,m}(1)		& 			&	\vdots		\\
h_{k,m}(L) 	&	\vdots			& 	\ddots	&	0		\\
0			&	h_{k,m}(L)		&			&	h_{k,m}(1)	\\
\vdots		&	\vdots			&	\ddots	&	\vdots		\\
0		 	&	0				&			&	h_{k,m}(L)	\\
\end{bmatrix}\nonumber
\end{IEEEeqnarray}
That is, $\HH_{k,m}$ is constructed with the CIR vector $\h_{k,m} = \left[ h_{k,m}(1),\ldots , h_{k,m}(L)\right]^T \in \C^L$, where $L$ is the finite CIR duration. We also define the channel matrix to user $k$ as $\HH_k = \left[ \HH_{k,1} \cdots \HH_{k,M} \right] \in \C^{B_r \times B_t M}$, i.e. the stacking of the channels matrices between all transmitter antennas and user $k$. Note that the received signal is spread in the time domain ($B_t$ transmitted symbols are spread across $B_r = B_t+L-1$ received samples). The CIR time samples $\{h_{k,m}(t)\}$ are zero-mean complex circularly-symmetric Gaussian random variables with diagonal covariance matrices $\Q_\h = \E\left\{\h_{k,m} \h_{k,m} ^H\right\} \in \C^{L \times L}$, $\forall k,m$. A common model for the diagonal elements of $\Q_\h$ (the channel power delay profile) is \cite{erceg2004} 
\begin{IEEEeqnarray}{rCl}
\label{eq_powerdelayprofile}
\left[ \Q_\h \right]_{ll} = \left( \frac{1-e^{\frac{t_s}{\sigma_\h}}}{1-e^{\frac{Lt_s}{\sigma_\h}}} \right) e^{-\frac{(l-1)t_s}{\sigma_\h}},
\end{IEEEeqnarray}
where $t_s$ is the sampling time, and $\sigma_\h$ is the mean channel delay spread. The factor in parenthesis in (\ref{eq_powerdelayprofile}) normalizes the channel power to satisfy the constraint $\Tr\left( \Q_\h \right) = 1$, $\forall m, k$. The diagonal structure of $\Q_\h$ ensures that the CIRs are uncorrelated across users, antennas, and time. This assumption is made in order to determine fundamental limits on the performance of frequency-selective MIMO broadcast channels, which are achieved under such uncorrelated scattering conditions.

\subsection{Receivers}
One of the main advantages provided by single-carrier frequency-selective channels over their multi-carrier counterparts is the reduced complexity at the receiver. We consider simple linear receiver structures, where $\G_k \in \C^{B \times B_r}$ represents a time-domain linear combiner at user $k$. In this work, we use two types of receivers. The first is a simple receiver that discards the first $\lceil (L+L_p-2)/2 \rceil$ and the last $\lfloor (L+L_p-2)/2 \rfloor$ time samples of the received block. This is the most common receiver in TR systems, since the discarded samples are ISI only \cite{viteri2015}. This filter has the form
\begin{IEEEeqnarray}{rCl}
\G_k = g_k \Gb = g_k 
\begin{bmatrix}
\ze 	&	\I_B	&	\ze 	\\
\end{bmatrix}, \nonumber
\end{IEEEeqnarray}
where $g_k \in \R^+$ represents an arbitrary gain control, $\Gb \triangleq \left[ \ze \ \  \I_B \ \, \ze \right]$ is the sample drop matrix, and $\ze$ is a zero matrix. We describe the second linear receiver in Section \ref{sec_jp}, where we exploit channel knowledge to improve the system performance.

The last component in the receiver signal in (\ref{eq_received}) is $\z_k \in \C^{B_r}$, which is the vector of time-domain noise samples. We assume $\z_k $ is a complex circularly-symmetric Gaussian random vector with covariance matrix $\Q_\z = \eta \I_{B_r}$, $\forall k$, where $\eta$ is the average noise power per sample. According to (\ref{eq_received}), the desired symbol block $\s_k$ is subject to a linear transformation induced by the matrix $\G_k \HH_{k} \PP_{k}$, with its diagonal elements representing the desired signal, while the off-diagonal elements correspond to ISI. IUI is determined by the matrices $\G_k \HH_{k} \PP_{k'}$ with $k' \neq k$. We define the desired signal, ISI, IUI, and noise power gains as
\begin{IEEEeqnarray}{rCl}
\label{eq_coefficients}
&& \alpha_{\dd,k} = \left\Vert (\G_k \HH_{k} \PP_{k}) \circ \I_B \right\Vert_F^2, \nonumber \\
&& \alpha_{\isi,k} = \left\Vert \G_k \HH_{k} \PP_{k} \right\Vert_F^2 - \alpha_{\dd,k} , \nonumber \\
&& \alpha_{\iui,k} = \sum_{k'\neq k} \rho_{k'} \left\Vert \G_k \HH_{k} \PP_{k'} \right\Vert_F^2 \nonumber\\
&& \alpha_{\nn,k} = \left\Vert \G_k \right\Vert_F^2, \nonumber
\end{IEEEeqnarray}
respectively, where $\circ$ denotes Hadamard product. Thus, the effective signal to interference plus noise ratio at receiver $k$ is
\begin{IEEEeqnarray}{rCl}
\label{eq_sinr}
\sinr_k = \frac{\rho_k \alpha_{\dd,k}}{\rho_k \alpha_{\isi,k}+\alpha_{\iui,k}+\eta \alpha_{\nn,k}}.
\end{IEEEeqnarray}
Note that, in frequency-selective MU-MIMO systems, both ISI and IUI are significant impairments for signal detection.

%% file: BDPrecoding.tex
\section{Block Diagonalization for Frequency-Selective Channels}
\label{sec_bd}
BD was first proposed for frequency-flat MU-MIMO channels in \cite{spencer2004}. The idea is to design a precoder such that the equivalent channel matrix $\HH\PP$ has a block diagonal structure. Thus, BD sets the IUI at every receiver to zero and the received signal in (\ref{eq_received}) has only the first and third terms. This allows a per-user precoder design since (\ref{eq_received}) depends only on the user index $k$. In the original formulation, BD is performed over a channel matrix with only spatial information between transmitter and receiver. However, the frequency-selective channel matrix comprises both space and time channel information. In this section, we analyze the particular structure of BD for frequency-selective channels, and propose three techniques to tackle its specific challenges. For the first two techniques, we assume a sample drop receiver $\G_k = \Gb$ and focus on the precoder design. We also assume perfect channel state information (CSI) at the transmitter. For the third technique we jointly design $\PP_k$ and $\G_k$ assuming CIS is also available at the receiver. From the received signal (\ref{eq_received}), IUI is set to zero when $\HH_k\PP_{k'} = \ze$, if $k \neq k'$. If we define the interference matrix for user $k$ as the stacking:
\begin{IEEEeqnarray}{rCl}
\label{eq_interfmatrix}
\Htil_k =
\begin{bmatrix}
\HH_1^T	&	\cdots	&	\HH_{k-1}^T	&	\HH_{k+1}^T	&	\cdots	&	\HH_K^T	\\
\end{bmatrix}^T,\nonumber
\end{IEEEeqnarray}
the condition for BD is $\Htil_k\PP_k = \ze, \ \forall k$, i.e. the columns of $\PP_k$ must lie in the null space of $\Htil_k$. Thus, as the first step to design the precoder $\PP_k$, we perform the singular value decomposition (SVD) of $\Htil_k$, in order to obtain a basis for $\nulls \left( \Htil_k \right)$. This SVD can be written as
\begin{IEEEeqnarray}{rCl}
\Htil_k = \Util_k \Sigtil_k \left[ \Vtil_k^{(1)} \ \, \Vtil_k^{(0)} \right]^H,\nonumber
\end{IEEEeqnarray}
where $\left[ \Vtil_k^{(1)} \ \, \Vtil_k^{(0)} \right]^H$ is the matrix formed with the right singular vectors of $\Htil_k \in \C^{B_r(K-1) \times B_tM}$. More specifically, the columns of $\Vtil_k^{(0)}$ form a basis for the null space of $\Htil_k$. Note that this matrix defined in (\ref{eq_interfmatrix}) is a column stacking of matrices taken from the set $\left\{ \HH_k \right\}$, so it is almost surely full (row or column) rank. Thus, unlike BD in frequency-flat channels, the dimension of $\Vtil_k^{(0)}$ in frequency-selective channels is known to be $B_t M \times B_v$, where $B_v = B_t M - B_r (K-1)$, independent of the propagation conditions. Hence, a BD precoder for the frequency-selective channel $\HH_k$ must have the form
\begin{IEEEeqnarray}{rCl}
\label{eq_generalbd}
\PP_k = \Vtil_k^{(0)}\Pb_k, \nonumber
\end{IEEEeqnarray}
where $\Pb_k \in \C^{B_v \times B}$ maps the transmitted block to user $k$ to the domain of $\Vtil_k^{(0)}$. Consequently, the search space for a BD precoder increases by using a larger number of antennas $M$ or reducing number of users $K$. The linear transformation $\G_k\HH_k\Vtil_k^{(0)}\Pb_k$ must be full rank so that the transmitted symbol block $\s_k$ can be recovered at the receiver. Therefore, both $\rank\left(\Vtil_k^{(0)}\right)\geq B$ and $\rank\left(\Pb_k\right) \geq B$ must hold, implying that $B_v \geq B$, and we can obtain the following conditions on the block size $B$ and the redundancy length $L_p$ for BD to be possible:
\begin{IEEEeqnarray}{rCl}
\label{eq_bdcondition2}
&& B \left( \frac{M-K}{M-K+1} \right) + L_p - 1 \geq \frac{(K-1)(L-1)}{M-K+1}, \nonumber\\
&& M \geq K. 
\end{IEEEeqnarray}
The first inequality can be used as a design criteria by either fixing $B$ or $L_p$, and then calculating the requirements on the other parameter. The second inequality states that the number of antennas must be greater than or equal to the number of users. The design problem then corresponds to finding the best matrix $\Pb_k$ to satisfy given performance optimization criteria for the desired signal transformation $\G_k\HH_k\Vtil_k^{(0)}\Pb_k$. An intuitive approach is to design $\Pb_k$ to provide some form of equalization (ISI mitigation), since IUI is already set to zero by using $\Vtil_k^{(0)}$. In the following, we propose two approaches to find $\Pb_k$, namely time-reversal-based BD and equalized BD, which use a simple receiver of the form $\G_k = \Gb$. We also present a third technique to jointly design $\PP_k$ and $\G_k$, using channel knowledge at the receiver. We present the solutions to the proposed optimization problems in the Appendices.

\subsection{Time-Reversal-Based Block Diagonalization}
TR beamforming is an emerging technique for SDMA over frequency-selective MU-MIMO channels. TR uses the complex-conjugate time-reversed CIR as a FIR filter at the transmitter, and yields space-time focusing of the signal at each receiver \cite{han2012,viteri2015}. In TR, the precoder $\PP_{m,k}$ is a (banded Toeplitz) convolution matrix constructed from the vector $\h_{k,m}^{\timer} = \left[ h_{k,m}^*(L), \ldots , h_{k,m}^*(1) \right]^T$ as
\begin{IEEEeqnarray}{rCl}
\PP_{m,k}^{\timer} =
\begin{bmatrix}
h_{k,m}^*(L)	&	0					&	\cdots	&	0				\\
\vdots			&	h_{k,m}^*(L)		& 			&	\vdots			\\
h_{k,m}^*(1) 	&	\vdots				& 	\ddots	&	0				\\
0				&	h_{k,m}^*(1)		&			&	h_{k,m}^*(L)	\\
\vdots			&	\vdots				&	\ddots	&	\vdots			\\
0		 		&	0					&			&	h_{k,m}^*(1)	\\
\end{bmatrix}, \nonumber
\end{IEEEeqnarray}
where the first factor ensures the precoder normalization. Note also that the redundancy is the same as the CIR length ($L_p = L$). We denote the TR precoder for user $k$ as
\begin{IEEEeqnarray}{rCl}
\Hbar_k = \left( B_t \sum_{m=1}^M \left\Vert \h_{k,m}^{\timer} \right\Vert_2^2 \right)^{-\frac{1}{2}}\left[ \PP_{1,k}^{\timer T} \cdots \PP_{M,k}^{\timer T} \right]^T \in \C^{M (B+L-1) \times B}. \nonumber
\end{IEEEeqnarray}
TR maximizes the desired signal power at the receiver by acting as a matched-filter, but its performance is limited by both ISI and IUI. We propose time-reversal based BD (TRBD) to take advantage of those properties of TR while eliminating IUI. This approach is similar to the frequency-domain approach in \cite{viteri2015}. The idea of TRBD is to obtain the closest precoder (in the minimum squared error sense) to the TR prefilter such that BD is achieved, which can be found by solving
\begin{IEEEeqnarray}{rCl}
\label{eq_trbdproblem}
\min_{\Pb_k} \left\Vert \Vtil_k^{(0)}\Pb_k - \Hbar_k \right\Vert_F^2,\quad \text{s.t. } \left\Vert \Vtil_k^{(0)}\Pb_k \right\Vert_F^2 = 1.
\end{IEEEeqnarray}
This problem has a closed-form solution (see Appendix \ref{appA}) such that the TRBD precoder is given by
\begin{IEEEeqnarray}{rCl}
\PP_k^{\timer} = \Vtil_k^{(0)} \frac{ \Vtil_k^{(0)H} \Hbar_k}{\left\Vert \Vtil_k^{(0)H} \Hbar_k \right\Vert_F} \in \C^{M(B+L-1) \times B},
\end{IEEEeqnarray}

\subsection{Equalized Block Diagonalization}
The performance of TR-based techniques is limited by ISI since TR pre-filters act only as partial equalizers: they maximize the desired signal power in (\ref{eq_received}) but they do not mitigate ISI explicitly. Henceforth, we propose a second strategy for precoder design, which aims to diagonalize the desired signal transformation, i.e. $\Gb\HH_k\PP_k \approx \I_B$. This design criteria is equivalent to maximize the desired signal to ISI power ratio. A complete diagonalization of the form $\Gb\HH_k\PP_k = \I_B$ is not attainable since an overdetermined system of linear equation results for the precoder. However, ISI can still be minimized by a least squares solution. In our particular BD model, the problem can be stated as
\begin{IEEEeqnarray}{rCl}
\label{eq_ebdproblem}
\min_{\Pb_k} \left\Vert \CC_k \Pb_k - \I_B \right\Vert_F^2,\quad \text{s.t. } \left\Vert \Vtil_k^{(0)}\Pb_k \right\Vert_F^2 = 1.
\end{IEEEeqnarray}
where $\CC_k = \Gb \HH_k \Vtil_k^{(0)}  \in \C^{B \times B_v}$. We refer to this approach as equalized block diagonalization (EBD). The solution for the precoder (see Appendix \ref{appB}) is
\begin{IEEEeqnarray}{rCl}
\label{eq_ebdprecoder}
\PP_k^{\eq} = \Vtil_k^{(0)} \left( \CC_k^H \CC_k + \mu_k \I_{B_c} \right)^{-1} \CC_k^H,
\end{IEEEeqnarray}
where $L_p > 0$ is arbitrarily chosen, $\mu_k \in \R$ is a Lagrange multiplier satisfying the first-order necessary condition
\begin{IEEEeqnarray}{rCl}
\label{eq_lagrangeebd}
\sum_{i=1}^{B_v}\frac{\lambda_{C_k,i}}{ \left( \lambda_{C_k,i}+\mu_k \right)^2 } = 1,
\end{IEEEeqnarray}
and $\{ \lambda_{C_k,i} \}_{i=1}^{B_v}$ is the set of eigenvalues of the positive definite matrix $\CC_k^H \CC_k$. The left-hand side in (\ref{eq_lagrangeebd}) is a monotonically decreasing function of $\mu_k$, so a unique solution can be easily found numerically by using a line search algorithm.

%% file: JointProcessing.tex
\subsection{Joint Transmitter/Receiver Processing in BD}
\label{sec_jp}
Both TRBD and EBD assume a sample drop receiver $\Gb$, but cannot eliminate ISI in the received signal. Thus, we propose a joint precoder/receiver design for BD when CSI is available at both the transmitter \emph{and} the receiver. We show that perfect equalization is possible using joint processing, such that both ISI and IUI are completely eliminated. The idea is to design both $\G_k$ and $\Pb_k$ such that $\G_k \HH_k \Vtil_k^{(0)} \Pb_k = \I_B$. We refer to this approach as joint processing block diagonalization (JPBD). We begin with the SVD of the equivalent block diagonalized channel, that is
\begin{IEEEeqnarray}{rCl}
\HH_k \Vtil_k^{(0)} = \U_k \Sig_k \V_k^H \in \C^{B_r \times B_v},
\end{IEEEeqnarray}
where $\U_k$ and $\V_k$ are unitary matrices, $\Sig_k = \diag \left( \sigma_{k,1}, \ldots , \sigma_{k,B_r} \right)$, and we assume $B_v \geq B_r$ so that the pseudoinverse of $\Sig_k$ satisfies $\Sig_k \Sig_k^+ = \I_{B_r}$. This assumption holds if
\begin{IEEEeqnarray}{rCl}
\label{eq_jprequiredb}
B_t \geq \frac{K(L-1)}{M-K} \quad \text{and} \quad M > K.
\end{IEEEeqnarray}
Note that (\ref{eq_jprequiredb}) is a stronger condition on the transmitter block length than (\ref{eq_bdcondition2}), viz. the number of transmit antennas must be strictly greater than the number of users. The system achieves a complete channel diagonalization if the precoder and receiver filter matrices are designed as $\PP_k^{\jp} = \Vtil_k^{(0)} \V_k \Sig_k^+  \Pu_k$ and $\G_k^{\jp}  = \Gu_k \U_k^H$, where $\Pu_k \in \C^{B_r \times B}$ projects the transmitted block of size $B$ to the received signal space of dimension $B_r = B + L + L_p - 2$ and $\Gu_k \in \C^{B \times B_r}$ reverses this operation. Using these matrices, the linear transformation corresponding to the desired signal in (\ref{eq_received}) is $\G_k^{\jp} \HH_k \PP_k^{\jp} = \Gu_k \Pu_k$. The final step in the design is to find the matrices $\Gu_k$ and $\Pu_k$ that satisfy $\Gu_k \Pu_k \propto \I_B$. A possible approach to this problem is to set $\Gu_k$ and $\Pu_k$ such that $\rank \left( \Gu_k \right)  =  B$ and $\Pu_k  =  \Gu_k^+ / \left\Vert \Sig_k^+ \Gu_k^+ \right\Vert_F$ (the precoder is normalized). The first condition ensures that $\Gu \Gu^+ = \I_B$, so ISI is completely eliminated. Using this approach the SINR at user $k$ is 
\begin{IEEEeqnarray}{rCl}
\label{eq_sinrjp1}
\sinr_k^{\jp} = \frac{\rho_k \alpha_{\dd,k}}{\eta \alpha_{\nn,k}} = \frac{\rho_k}{\eta} \left\Vert \Sig_k^+ \Gu_k^+ \right\Vert_F^{-2} \left\Vert \Gu_k \right\Vert_F^{-2}.
\end{IEEEeqnarray}
Hence, we can select the matrix $\Gu_k$ that maximizes the SNR by solving
\begin{IEEEeqnarray}{rCl}
\min_{\Gu_k} \left\Vert \Sig_k^+ \Gu_k^+ \right\Vert_F^2 \left\Vert \Gu_k \right\Vert_F^2.
\end{IEEEeqnarray}
Optimality conditions for this problem lead to a nonlinear matrix equation with no general closed-form solution (see Appendix \ref{appC}). However, if we assume that $\Gu_k$ is a rectangular diagonal matrix with real positive entries, a closed-form solution to this problem exists and is given by
\begin{IEEEeqnarray}{rCl}
\label{eq_gudiag}
\left[ \Gu_k \right]_{ii} = \sqrt{\frac{1}{\sigma_i}},
\end{IEEEeqnarray}
where $\sigma_{k,i}$ is the $i$-th singular value of $\HH_k \Vtil_k^{(0)}$. Hence, the precoder and receiver filter in JPBD are
\begin{IEEEeqnarray}{rCl}
\label{eq_jpsolution}
\PP_k^{\jp} & = & \Vtil_k^{(0)} \V_k \frac{ \Sig_k^+  \Gu_k^+ } { \left\Vert \Sig_k^+ \Gu_k^+ \right\Vert_F }, \\
\G_k^{\jp}  & =&  \Gu_k \U_k^H.
\end{IEEEeqnarray}
The JPBD precoder and receiver filter resemble the conventional BD solution in \cite{spencer2004} by using: \emph{i}) the interference suppression provided by $\Vtil_k^{(0)}$, \emph{ii}) the eigenbeamformers $\U_k$ and $\V_k$, which share the role of eliminating ISI, and \emph{iii}) the amplitude equalizers $\Sig_k^+  \Gu_k^+$ and $\Gu_k$, which ensure that all symbols in the received block have the same average power. Note that $\HH_k \Vtil_k^{(0)}$ has $B_r$ singular values, but only $B$ of them are used to calculate the JPBD solution. The influence of these singular values on the performance of JPBD is analyzed in Section \ref{sec_analysis}. In addition, using (\ref{eq_sinrjp1}) and (\ref{eq_gudiag}), the SINR in terms of the singular values of $\HH_k \Vtil_k^{(0)}$ is 
\begin{IEEEeqnarray}{rCl}
\label{eq_sinrjp}
\sinr_k^{\jp} = \frac{\rho_k}{\eta} \left( \sum_{i=1}^B \frac{1}{\sigma_{k,i}} \right)^{-2}.
\end{IEEEeqnarray}

%% file: PowerAllocation.tex
\subsection{Power Allocation for Sum-Rate Maximization}
\label{sec_poweralloc}
In the previous section, we presented three linear processing techniques for the frequency-selective MIMO broadcast channel. Both TRBD and EBD do not eliminate ISI in the received signal, so conventional waterfilling \cite{goldsmith2003} cannot be applied for power allocation. Thus, in this section we propose a power allocation scheme for sum-rate maximization in TRBD and EBD, which takes into account ISI in the received signal. Maximizing the sum-rate in the downlink subject to a maximum power constraint can be stated as
\begin{IEEEeqnarray}{rCl}
\label{eq_powerrateproblem}
\hspace*{-20pt}\max_{\rhob} \sum_{k=1}^K \log_2 \left( 1 + \sinr_{k} \right), \text{ s.t. }  \ \Vert \rhob \Vert_1 \leq P_{\max}, \ \rhob \geq 0,
\end{IEEEeqnarray}
where $\rhob = \left[ \rho_1 , \ldots , \rho_K \right]^T$ is the vector of transmitted powers, and $\Vert \cdot \Vert_1$ denotes $\ell_1$ vector norm.
Using the Lagrange multiplier method (see Appendix \ref{appD}), the optimal power allocation in this case is
\begin{IEEEeqnarray}{rCl}
\rho_k & = & \frac{\sqrt{\eta \alpha_{\dd,k}^2 \alpha_{\nn,k}^2 + \frac{4 \eta \alpha_{\dd,k} \alpha_{\isi,k} \alpha_{\nn,k}}{\lambda \ln(2)}(\alpha_{\dd,k}+\alpha_{\isi,k})} } {2 \alpha_{\isi,k} (\alpha_{\dd,k}+\alpha_{\isi,k})} \nonumber \\
&& - \frac{\eta \alpha_{\nn,k} (\alpha_{\dd,k}+2\alpha_{\isi,k})}{2 \alpha_{\isi,k} (\alpha_{\dd,k}+\alpha_{\isi,k})},
\end{IEEEeqnarray}
where $\lambda$ is a Lagrange multiplier satisfying
\begin{IEEEeqnarray}{rCl}
\label{eq_lagrangepower}
\hspace*{-20pt} \sum_{k=1}^K \frac{\sqrt{\eta^2 \alpha_{\dd,k}^2 \alpha_{\nn,k}^2 + \frac{4 \eta \alpha_{\dd,k} \alpha_{\isi,k} \alpha_{\nn,k}}{\lambda \ln(2)}(\alpha_{\dd,k}+\alpha_{\isi,k})}}{2 \alpha_{\isi,k} (\alpha_{\dd,k}+\alpha_{\isi,k})} && \nonumber \\
\hspace*{-20pt} - \sum_{k=1}^K \frac{\eta \alpha_{\nn,k} (\alpha_{\dd,k}+2\alpha_{\isi,k})}{2 \alpha_{\isi,k} (\alpha_{\dd,k}+\alpha_{\isi,k})} & = & P_{\max}.
\end{IEEEeqnarray}
Note that the left hand side in (\ref{eq_lagrangepower}) is a monotonically decreasing function of $\lambda$, so its unique value satisfying the constraint can be found by using a line search algorithm. This search should be limited to the interval $0 < \lambda \leq \min_k (\alpha_{\dd,k}/ [ \alpha_{\nn,k} \ln(2)])$ so that $\rhob \geq 0$ holds. In the case of JPBD, since ISI is completely eliminated, conventional waterfilling can be applied for power allocation using the signal to noise ratio in (\ref{eq_sinrjp}).

%% file: Analysis.tex
\section{Performance Analysis of Frequency-Selective BD techniques}
\label{sec_analysis}

In this section, we analyze the performance of BD methods for frequency-selective channels under different SNR regimes. For high SNR, the system is characterized by $\rho_k / \eta \to \infty$, $\forall k$, which implies $P_{\max}/\eta \to \infty$ given the power constraint $\sum_k \rho_k = P_{\max}$. In this case, we analyze the diversity and the multiplexing gains for each BD method. When the system operates at low SNR, the term associated to noise dominates the denominator in (\ref{eq_sinr}), i.e. $\eta \alpha_{\nn,k} \gg \rho_k \alpha_{\isi,k}$, and we obtain a technique-independent upper bound for the SINR.

\subsection{Multiplexing Gain}
Assuming the receivers treat interference as Gaussian noise, we define the ergodic achievable rate for user $k$ as
\begin{IEEEeqnarray}{rCl}
\label{eq_rateperuser}
R_k \left( \sinr_k \right)  & = &  \left( \frac{B}{B+L+L_p-2} \right) \E \left\{ \log_2 \left( 1 + \sinr_k \right) \right\},\nonumber \\
& = & \left( \frac{B}{B+L+L_p-2} \right) \E \left\{ \log_2 \left( 1 + \frac{\rho_k \alpha_{\dd,k}}{\rho_k \alpha_{\isi,k} + \eta \alpha_{\nn,k}} \right) \right\}, \nonumber \\
\end{IEEEeqnarray}
where the factor outside the expectation accounts for the guard interval, the expectation is taken over the channel matrix $\HH$, and $\sinr_k$ is given in (\ref{eq_sinr}) with $\alpha_{\iui,k} = 0$ since any BD technique eliminates IUI. The multiplexing gain for user $k$ is defined as
\begin{IEEEeqnarray}{rCl}
\label{eq_multiplexinggain}
r_k = \lim_{\frac{\rho_k}{\eta} \to \infty}  \frac{R_k\left( \sinr_k \right)}{\log_2 \left( \frac{\rho_k}{\eta} \right) },
\end{IEEEeqnarray}
and the \emph{system multiplexing gain} is
\begin{IEEEeqnarray}{rCl}
\label{eq_systemmultiplexinggain}
r = \sum_{k=1}^K  r_k.
\end{IEEEeqnarray}
Thus, $r$ is the slope in the achievable sum-rate $R=\sum_k R_k \left( \sinr_k \right)$ at high SNR when plotted against $P_{\max}/\eta$ (since $P_{\max}/\eta \to \infty$ implies $\rho_k/\eta \to \infty,\,\forall k$). Note that, if $\alpha_{\isi,k} \neq 0$, $\sinr_k \to \alpha_{\dd,k}/\alpha_{\isi,k}$ when $\rho_k/\eta \to \infty$. Consequently, since TRBD and EBD cannot (completely) eliminate ISI, their system multiplexing gains are
\begin{IEEEeqnarray}{rCl}
r^{\timer} = r^{\eq} = 0, \nonumber
\end{IEEEeqnarray}
respectively. In contrast, JPBD eliminates ISI and using L'H\^opital's rule with $\alpha_{\isi,k} = 0$ on (\ref{eq_rateperuser})-(\ref{eq_systemmultiplexinggain}), JPBD achieves a system multiplexing gain
\begin{IEEEeqnarray}{rCl}
\label{eq_jpmultiplexinggain}
r^{\jp} =  \frac{BK}{B+L+L_p-2},
\end{IEEEeqnarray}
where we have assumed that the channels for different users have the same statistics. Note that $\lim_{B \to \infty} r^{\jp} = K$, i.e. JPBD has full diversity gain (equal to the number of users) when the transmitted block size goes to infinity. Thus, JPBD outperforms other techniques in the high SNR regime.

\subsection{Diversity Gain}
The diversity gain for user $k$ is defined as
\begin{IEEEeqnarray}{rCl}
\label{eq_diversitygain}
d_k = - \lim_{\frac{\rho_k}{\eta} \to \infty}  \frac{ \E \left\{ \log \left[  P_e (\sinr_k)  \right] \right\}  }{\log \left( \frac{\rho_k}{\eta} \right) }, \nonumber
\end{IEEEeqnarray}
where $ P_e (\sinr_k)$ is the probability of error at user $k$. Assume that the symbols in $\s_k$ are taken from a QAM constellation. Then, the error probability at high SNR is approximately \cite[Sec. 9.1.2]{tse2005}
\begin{IEEEeqnarray}{rCl}
P_e (\sinr_k) \approx \frac{1}{\sinr_k^{1 - r_k }}, \nonumber
\end{IEEEeqnarray}
which assumes the QAM rate increases continuously with SNR (this cannot be attained in practice, where discrete modulation orders are used). The diversity gain for user $k$ is then
\begin{IEEEeqnarray}{rCl}
\label{eq_diversitygain2}
d_k = (1 - r_k)  \lim_{\frac{\rho_k}{\eta} \to \infty} \frac{ \E \left\{ \log \left( \sinr_k \right) \right\}  }{\log \left( \frac{\rho_k}{\eta} \right) }.
\end{IEEEeqnarray}
The fact that $\sinr_k \to \alpha_{\dd,k}/\alpha_{\isi,k}$ if $\alpha_{\isi,k} \neq 0$ implies that the diversity gain for TRBD and EBD is 
\begin{IEEEeqnarray}{rCl}
d_k^{\timer} = d_k^{\eq} = 0,
\end{IEEEeqnarray}
respectively. In contrast, replacing (\ref{eq_sinrjp}) into (\ref{eq_diversitygain2}) gives the following diversity gain for JPBD
\begin{IEEEeqnarray}{rCl}
\label{eq_jpdiversitygain}
d_k^{\jp} = 1-r_k^{\jp} = \frac{L-L_p-2}{B+L+L_p-2}.
\end{IEEEeqnarray}
Thus, the diversity-multiplexing tradeoff is clearly observed \cite{zheng2003,tse2005}. According to (\ref{eq_jpmultiplexinggain}) and (\ref{eq_jpdiversitygain}), for a fixed block length $B$ a larger channel delay spread $L$ or a larger precoder redundancy $L_p$ improve the diversity gain but deteriorate the multiplexing gain. In contrast, for fixed $L$ and $L_p$, a larger block length improves the multiplexing gain but deteriorates the diversity gain.

\subsection{Low SNR Characterization}
Now, we derive a bound for the SINR at low SNR (i.e. $\eta \alpha_{\nn,k} \gg \rho_k \alpha_{\isi,k}$), and demonstrate that it is proportional to the the number of transmit antennas $M$ and the transmitted block length $B_t$. We assume the best case scenario where the equalization provided by any technique is such that $\alpha_{\isi,k} \approx 0$ and $\alpha_{\dd,k} = \left\Vert \G_k \HH_k \PP_k \right\Vert_F^2$. Under those conditions, the SINR is
\begin{IEEEeqnarray}{rCl}
\label{eq_sinrbound}
\sinr_k = \frac{\rho_k \left\Vert \G_k \HH_k \PP_k \right\Vert_F^2}{\eta \left\Vert \G_k \right\Vert_F^2} \leq \frac{\rho_k}{\eta}\left\Vert \HH_k \right\Vert_F^2,
\end{IEEEeqnarray}
where we used the submultiplicative property of Frobenius norms ($\Vert \A\B \Vert_F \leq \Vert \A \Vert_F \Vert \B \Vert_F$ for any matrices $\A$ and $\B$) \cite{laub2005}, and the precoder normalization $\left\Vert \PP_k \right\Vert_F^2 = 1$. Taking the expectation of (\ref{eq_sinrbound}) with respect to the channel yields
\begin{IEEEeqnarray}{rCl}
\label{eq_sinrbound2}
\E \left\{ \sinr_k \right\} & \leq & \frac{\rho_k}{\eta} M B_t.
\end{IEEEeqnarray}
Henceforth, the number of antennas on the frequency-selective MU-MIMO downlink provides a multiplicative gain on the low SNR regime, rather than the conventional improvement on the high SNR diversity and multiplexing gains of the frequency-flat case.

%% file: Results.tex
\section{Numerical Results and Discussion}
\begin{figure*}[!t]
\centering
\includegraphics[width=\columnwidth]{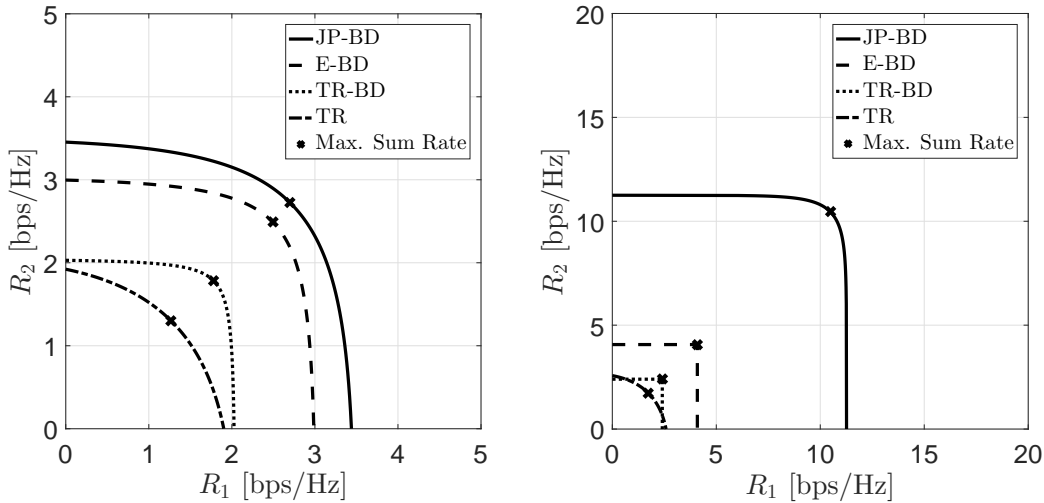}
\caption{Achievable rate regions of the proposed techniques with $P_{\max}/\eta = 20$ dB (left) and $50$ dB. The system has $M=8$ antennas.}
\label{fig_resultsrr}
\end{figure*}
\begin{figure*}[!t]
\centering
\includegraphics[width=\columnwidth]{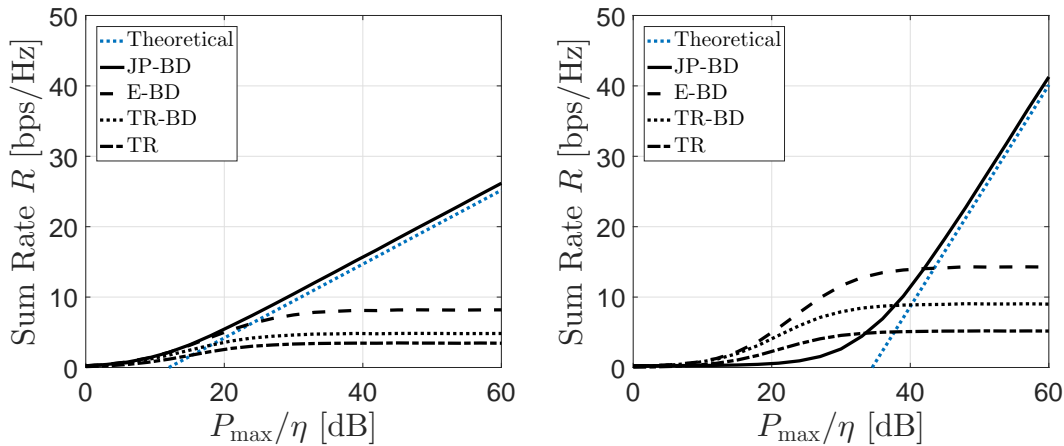}
\caption{Achievable sum rate for $K = 2$ (left), and $K=6$ users for $M=8$ antennas. The theoretical reference is a line with a slope equal to the multiplexing gain.}
\label{fig_resultssumratevspmax}
\end{figure*}
We performed extensive simulations of the three proposed BD techniques for frequency-selective channels using parameters as shown in Table \ref{tab_simparam} (unless indicated explicitly in each figure). We selected these values to approximate those of common WLAN channel models such as \cite{erceg2004}, and we assume the system operates over a 100 MHz bandwidth in a typical indoor scenario. Each random channel matrix realization was generated to match the model described in Section \ref{sec_channel}.
\begin{table}[!t]
\centering
\caption{Simulation Parameters}
\label{tab_simparam}
\begin{tabular}{lc}
\hline
\hline
\textbf{Parameter}						&	\textbf{Value}	\\
\hline
Mean delay spread ($\sigma_\h$)			&	15 ns			\\
Sampling time ($t_s$)					&	10 ns			\\
Block length ($B$)						&	30 symbols		\\
CIR duration ($L$)						&	9 samples		\\
Precoder redundancy$^\dagger$ ($L_p$)	&	1 sample		\\
Number of transmit antennas ($M$)		&	8				\\
Number of channel realizations			&	$10^3$			\\
\hline
\hline
\multicolumn{2}{l}{$^\dagger$ For EBD and JPBD. $L_p=1$ implies that the precoder does}\\
\multicolumn{2}{l}{\hspace*{7pt}not add time-domain redundancy. TRBD uses $L=L_p$.}\\
\hline
\hline
\end{tabular}
\end{table}

\subsection{Achievable Rate Regions}
Fig. \ref{fig_resultsrr} shows the limits of the achievable rate region for $K=2$ users under the power constraint $\rho_1+\rho_2 = P_{\max}$. The plot shows that TR, TRBD, and EBD improve slightly with a higher $P_{\max}/\eta$, since they are limited by ISI (as well as IUI in TR) and not by noise. JPBD capacity region expands when increasing $P_{\max}/\eta$ since it eliminates ISI and IUI completely. The achievable rate regions are close to squared in all BD techniques given that IUI is set to zero, which implies that increasing the transmitted power to a given user does not increase interference to the others.

\subsection{Achievable Sum Rate and Multiplexing Gain}
Fig. \ref{fig_resultssumratevspmax} (left and center) shows the maximum achievable sum rate as a function of $P_{\max}/\eta$ (using the power allocation scheme described in Section \ref{sec_poweralloc}). The figure shows that TR, TRBD, and EBD have a bound on the maximum sum rate when $P_{\max}/\eta \to \infty$ since they do not eliminate ISI completely (this corroborates the fact that their multiplexing gain is $r = 0$). It is also observed that JPBD has the best performance at high SNR and the simulated multiplexing gain shows good agreement with the theoretical results. Note that, when the number of users increases, higher SNR is required to achieve the same rate since less power is allocated per user.

\subsection{Bit Error Rate and Diversity Gain}
We analyze the average bit error rate (BER) per user performance of the proposed methods with the transmission of $10^6$ bits using QAM constellations of different orders. Fig. \ref{fig_resultsber} shows the BER with different number of antennas and different modulation orders. An approximate 6 dB gain is observed on the required $P_{\max}/\eta$ for JPBD when doubling the number of antennas, which is consistent with the bound in (\ref{eq_sinrbound2}) for two users (it translates to a 3 dB gain on $\rho_k/\eta$ for each user). It is also clear that TR, TRBD, and EBD cannot eliminate ISI, inducing a lower bound on the BER at high SNR. However, ISI can be mitigated by using a larger number of antennas, so a lower BER at high SNR is observed when increasing $M$. This characteristic of TR based systems has been also observed in other works \cite{viteri2015}. Fig. \ref{fig_resultsber} (right) shows the JPBD performance when increasing the QAM constellation size. Note that the diversity gain in (\ref{eq_jpdiversitygain}) assumes that the rate (constellation size) increases continuously with SNR, so $d_k^{\jp}$ gives a bound on the BER slope for increasing modulation order at high SNR. Thus, the diversity gain slope is better observed when the modulation order is increased with the SNR, e.g., Fig. \ref{fig_resultsber} (right) shows an adaptive-rate modulation where the modulation rate is $2^{R_{\qam}}$ and $R_{\qam}$ is the largest even integer smaller than or equal to $r_k \log_2 \left( P_{\max} / \eta \right)$ (this ensures a rectangular QAM constellation if $R_{\qam} \geq 4$). This adaptive modulation scheme shows good agreement with the diversity gain, according to the plot.

\subsection{Impact of the Number of Users}
Fig. \ref{fig_resultssingval} (left) shows the maximum achievable sum rate as a function of the number of users $K$, with all other system parameters kept constant. We used the power allocation in Section \ref{sec_poweralloc}. The figure shows that JPBD has the best performance again, followed by EBD, and TRBD. The sum rate in JPBD increases linearly until the number of users approaches the number of antennas (16 in this example) and then drops markedly when $L_p=1$ (no time-domain redundancy is added at the precoder). This behavior is caused by  the SINR dependence on the first $B$ singular values of $\HH_k\Vtil_k^{(0)}$ as given by (\ref{eq_sinrjp}). As discussed in Section \ref{sec_bd}, $\HH_k\Vtil_k^{(0)} \in \C^{B_r \times B_v}$ has $B_r$ non-zero singular values. Thus, a smaller $B_v = B_t M -B_r(K-1)$ (caused by increasing number of users), decreases the amplitude of those singular values and also the SINR in JPBD. A practical solution to this problem is to increase the precoder redundancy $L_p$, which increases both $B_v$ and the SINR enabling an almost linear growth in the sum rate when the number of users approaches the number of antennas. We observe this effect in Fig. \ref{fig_resultssingval} (center and right). However, increasing $L_p$ has a small impact on the sum rate when the number of users is low compared to the number of antennas.

%% file: Conclusions.tex
\section{Conclusion}

We explored the generalization of BD precoding techniques, originally proposed for frequency-flat MIMO broadcast channels, to the frequency-selective case. Such generalization is not straightforward since the channel matrix has a space-time structure constructed from the channel impulse responses. We derived the conditions under which BD is feasible for block transmissions in frequency-selective MIMO broadcast channels: the transmitted block length should be sufficiently large and the number of transmit antennas should be greater than or equal to the number of users (see inequality (\ref{eq_bdcondition2})).

Even though any BD eliminates IUI, frequency selectivity induces ISI in the received signal. Thus, we proposed three approaches to mitigate or suppress ISI. The first approach, TRBD, finds the BD precoder matrix which is closest (in the minimum squared error sense) to the TR pre-filter; although it improves the performance of conventional TR, it is still limited by ISI. EBD is the second approach, which explicitly minimizes ISI using an equalizer at the transmitter; EBD outperforms TR based solutions but cannot suppress ISI completely. Moreover, we showed that any precoding-only scheme which do not eliminate ISI has zero diversity and multiplexing gains (their achievable sum rates are bounded at high SNR). Thus, we propose a joint transmitter/receiver design called JPBD, which is based on the SVD of the equivalent block-diagonalized channel. We demonstrated that, for an infinite block length, JPBD achieves full multiplexing gain (equal to the number of users). We showed that the diversity gain in JPBD improves with larger channel delay spread or larger time-domain precoder redundancy, but decreases with larger block length $B$ (see eq. (\ref{eq_jpdiversitygain})).

Extensive numerical simulations show that all the proposed BD solutions for frequency-selective MIMO broadcast channels outperform conventional TR beamforming. Moreover, numerical results show good agreement with the theoretical results derived in this paper. We also examined the performance of each technique under different operation parameters, e.g. number of antennas, number of users, block length, and precoder redundancy.

\begin{figure*}[!t]
\centering
\subfloat{\includegraphics[width=0.5\columnwidth]{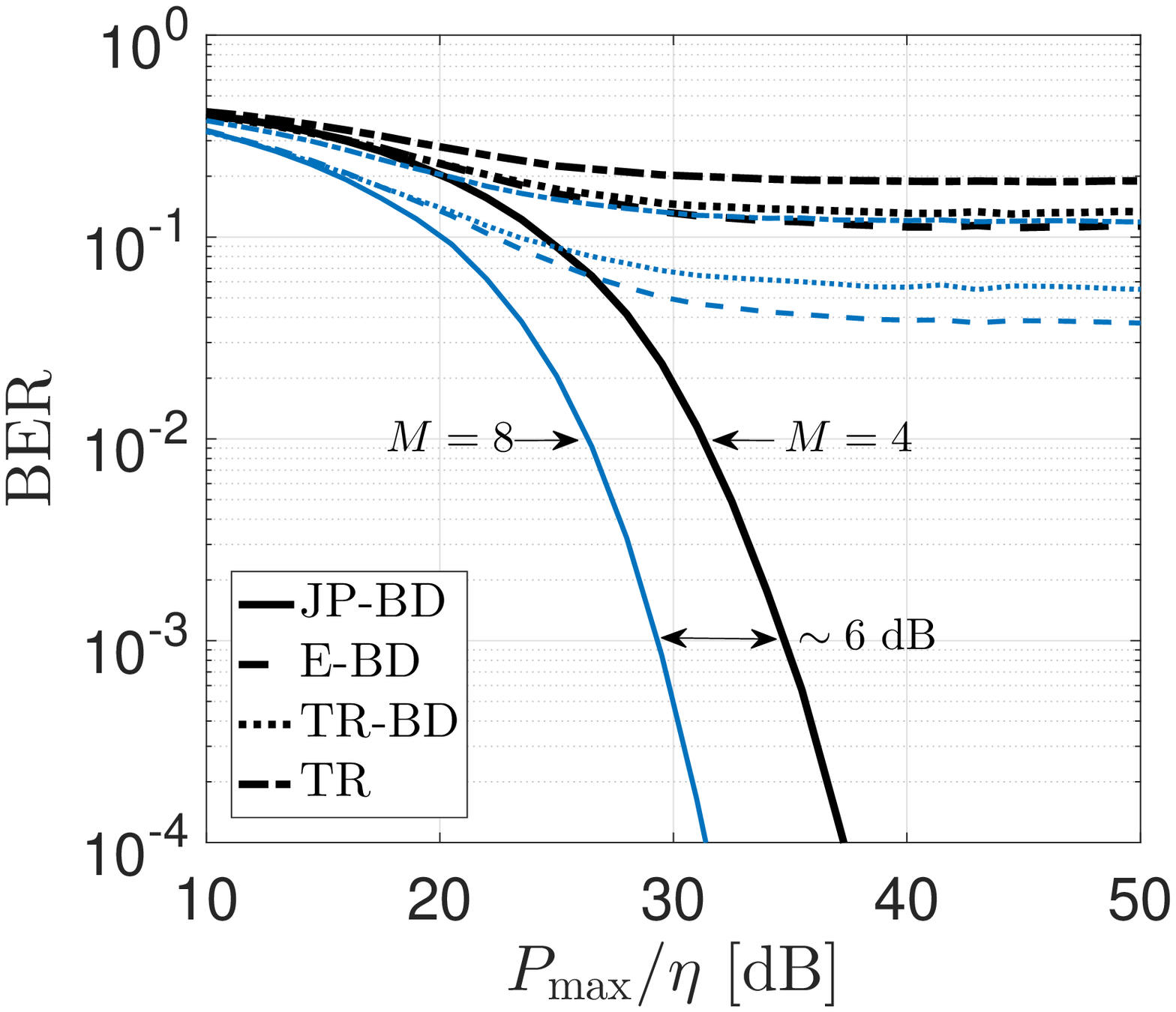}}
\subfloat{\includegraphics[width=0.5\columnwidth]{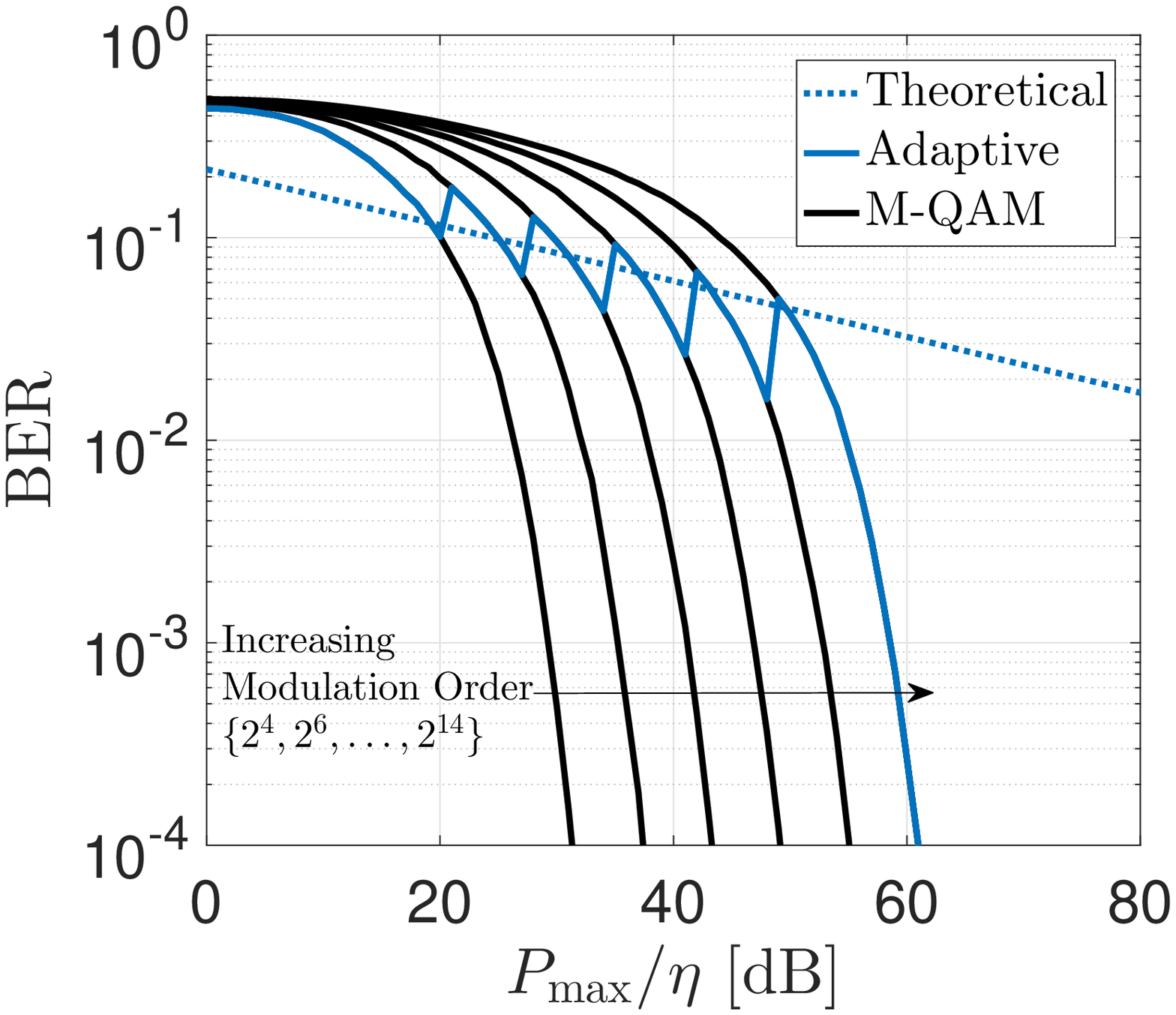}}
\caption{Bit error rate performance with $K = 2$ users and $L_p=1$ (no time-domain redundancy). On the left, 16-QAM BER with $B=100$ symbols, different number of antennas and techniques. On the right, JPBD BER performance with different QAM orders, adaptive modulation rate, and the theoretical reference (a line with a slope equal to the diversity gain and $B=50$ symbols).}
\label{fig_resultsber}
\end{figure*}
\begin{figure*}[!t]
\centering
\subfloat{\includegraphics[width=0.33\columnwidth]{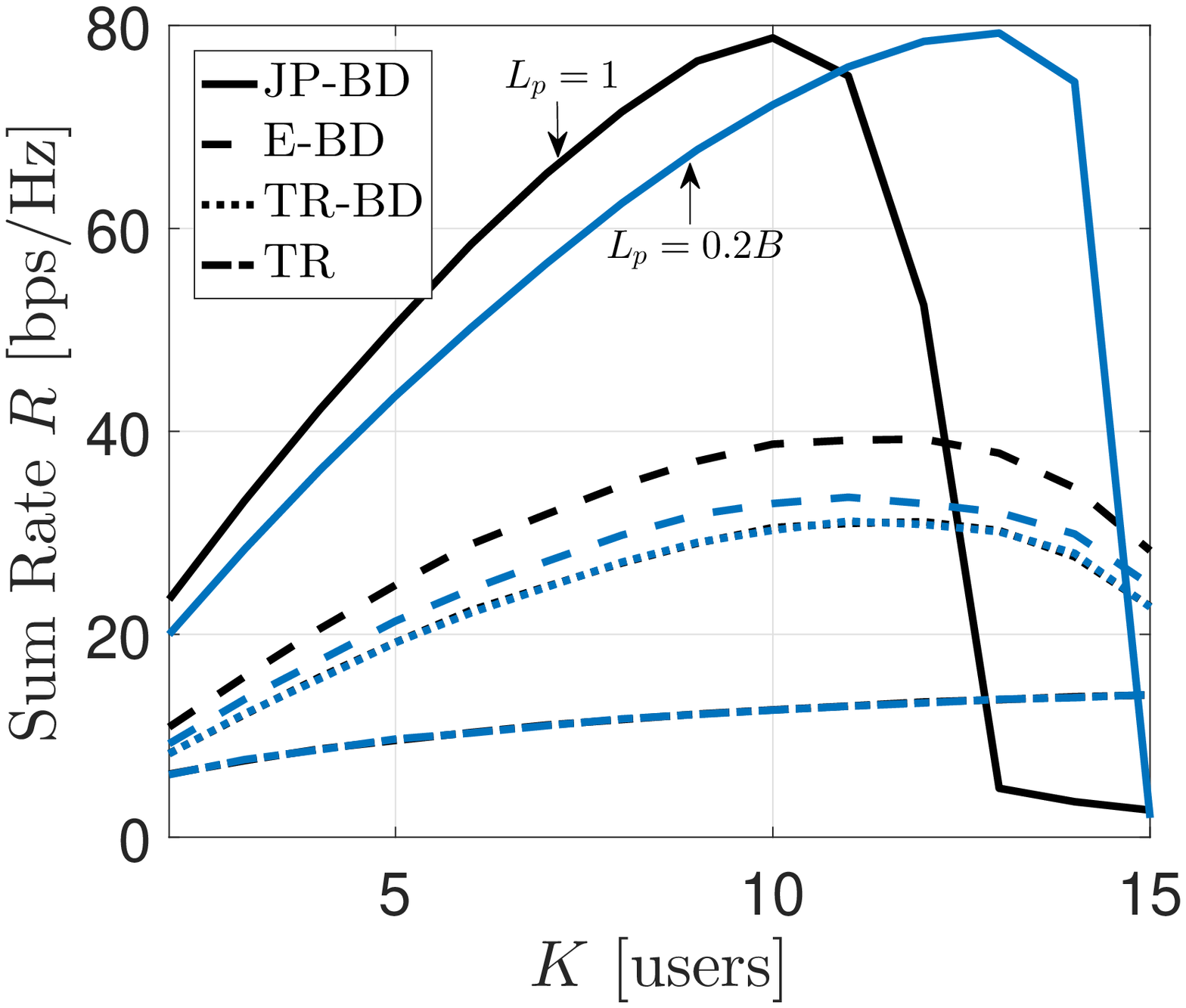}}
\subfloat{\includegraphics[width=0.33\columnwidth]{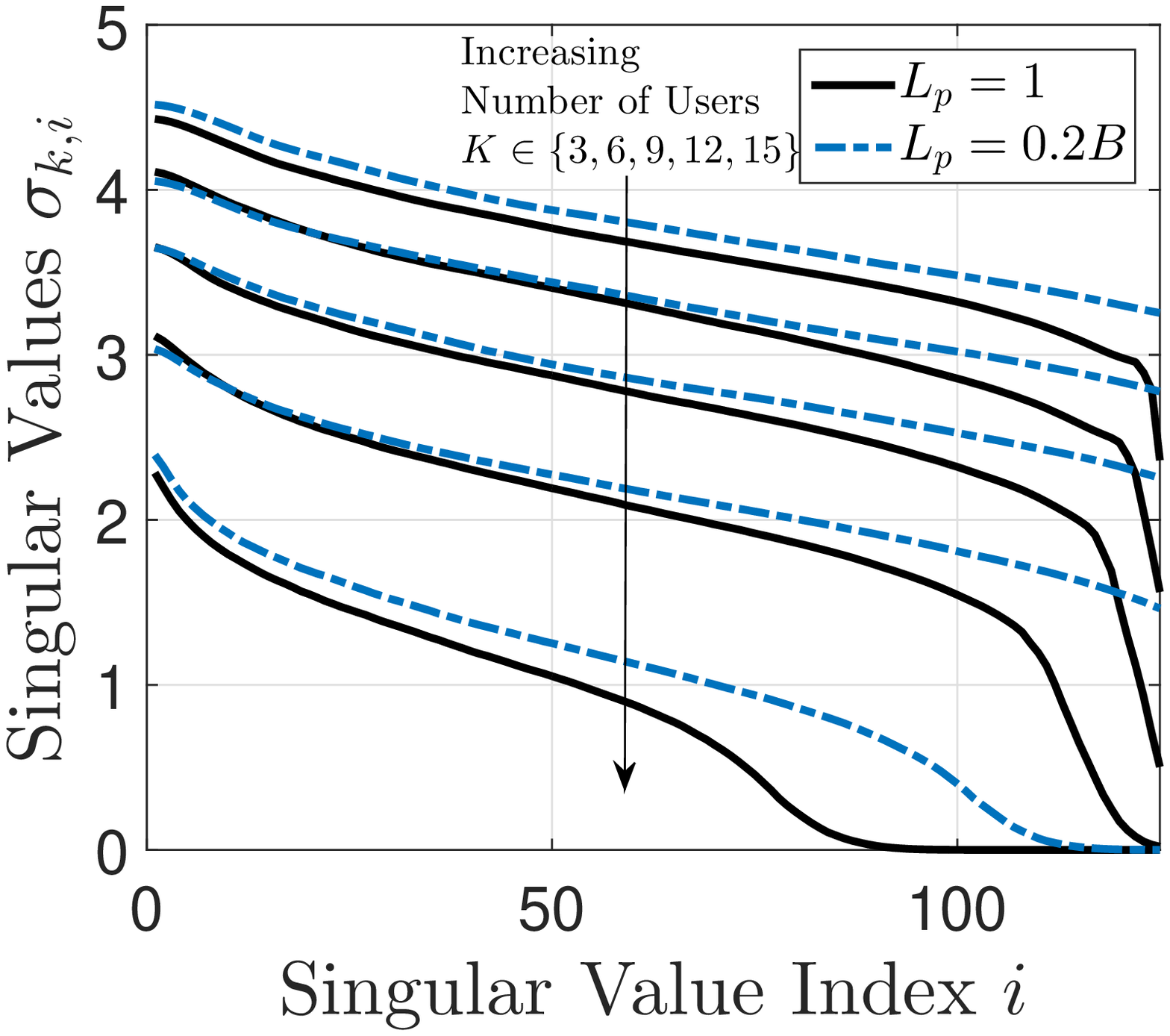}}
\subfloat{\includegraphics[width=0.33\columnwidth]{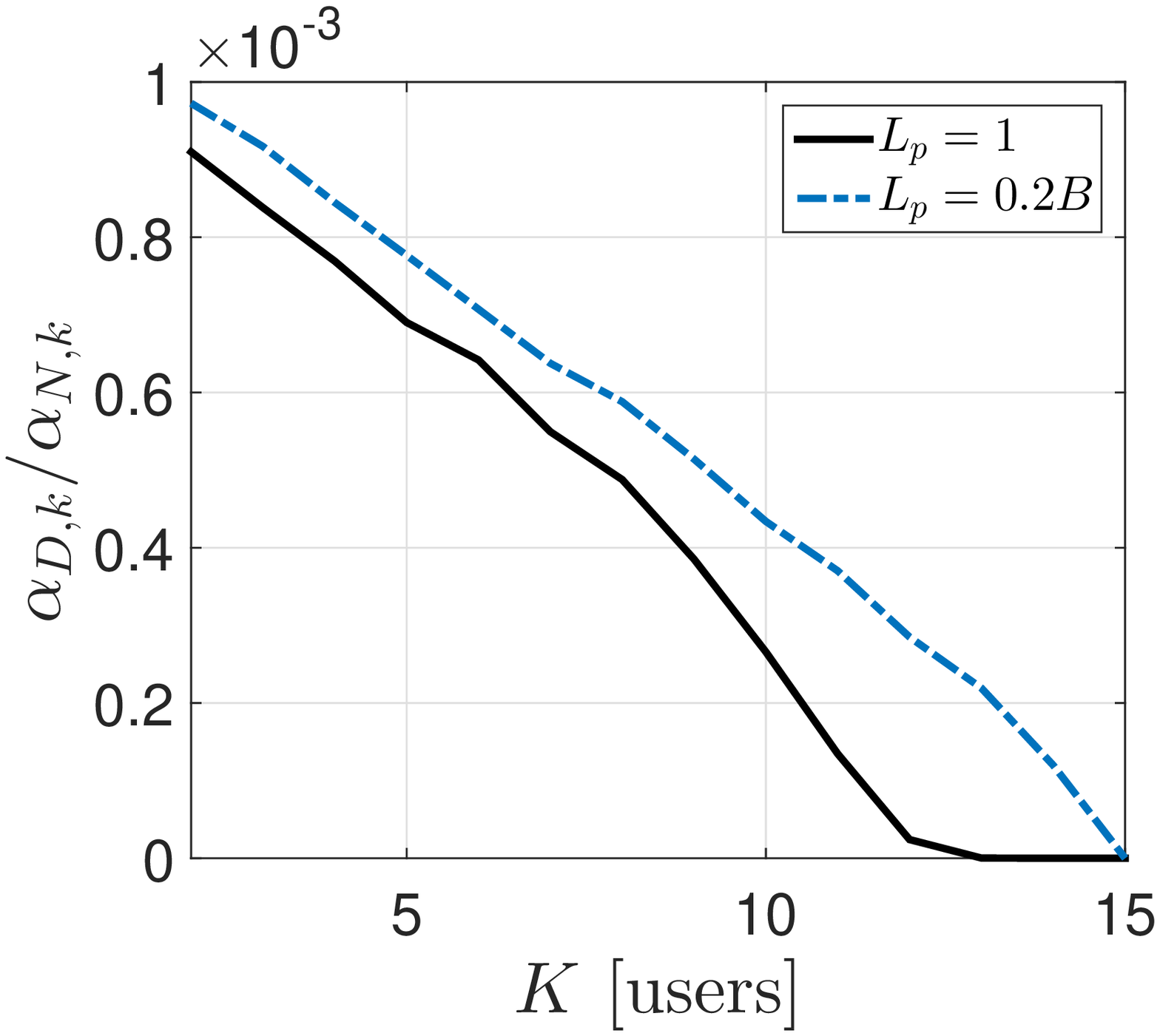}}
\caption{System analysis when the number of users increases with fixed $B = 80$ and $M=10$. (Left) Maximum achievable sum rate as a function of $K$. (Center) Singular values of $\HH_k \Vtil_k^{(0)}$. (Right) SINR coefficient for JPBD $\left(\alpha_{\dd,k}/\alpha_{\nn,k}\right)$.}
\label{fig_resultssingval}
\end{figure*}

%% file: Appendices.tex
\appendices

\section{TRBD Precoder Solution}
\label{appA}
We obtain the TRBD precoder design by solving
\begin{IEEEeqnarray}{rCl}
\label{eq_trbdproblemapp}
\min_{\Pb_k} \left\Vert \Vtil_k^{(0)}\Pb_k - \Hbar_k \right\Vert_F^2,\quad \text{s.t. } \left\Vert \Vtil_k^{(0)}\Pb_k \right\Vert_F^2 = 1,
\end{IEEEeqnarray}
The Lagrangian of (\ref{eq_trbdproblemapp}) is
\begin{IEEEeqnarray}{rCl}
&& \lag_{\timer}\left(\Pb_k , \lambda_k \right) \nonumber \\
&& = \left\Vert \Vtil_k^{(0)}\Pb_k - \Hbar_k \right\Vert_F^2 + \lambda_k \left( \left\Vert \Vtil_k^{(0)}\Pb_k \right\Vert_F^2 - 1 \right) \nonumber \\
&& = \Tr \left( \Vtil_k^{(0)}\Pb_k \Pb_k^H \Vtil_k^{(0)H} \right) - \Tr \left( \Vtil_k^{(0)}\Pb_k \Hbar_k^H \right) \nonumber \\
&& \quad - \Tr \left( \Hbar_k \Pb_k^H \Vtil_k^{(0)H} \right) + \lambda_k \left[ \Tr \left( \Vtil_k^{(0)}\Pb_k \Pb_k^H \Vtil_k^{(0)H} \right)  - 1 \right] \nonumber \\
&& = \Tr \left( \Pb_k \Pb_k^H  \right) - \Tr \left( \Vtil_k^{(0)}\Pb_k \Hbar_k^H \right) \nonumber \\
&& \quad - \Tr \left( \Hbar_k \Pb_k^H \Vtil_k^{(0)H} \right) + \lambda_k \left[ \Tr \left( \Pb_k \Pb_k^H  \right)  - 1 \right], \nonumber 
\end{IEEEeqnarray}
where we used the cyclic permutation invariance of the trace, and the fact that $ \Vtil_k^{(0)H} \Vtil_k^{(0)} = \I_{B_v}$ (the columns of $\Vtil_k^{(0)}$ are orthonormal). $\lambda_k \in \R$ is a Lagrange multiplier. Taking the Lagrangian derivative with respect to $\Pb_k$ yields the Karush-Kuhn-Tucker (KKT) condition \cite{bertsekas1999}
\begin{IEEEeqnarray}{rCl}
\label{eq_fonctr}
&& \frac{\partial \lag_{\timer}\left(\Pb_k , \lambda_k \right)}{\partial \Pb_k} =  \Pb_k^* - \Vtil_k^{(0)T} \Hbar_k^* +  \lambda_k \Pb_k^* = \ze.
\end{IEEEeqnarray}
Using the complex-conjugate of (\ref{eq_fonctr}) and applying the constraint $\left\Vert \Vtil_k^{(0)}\Pb_k \right\Vert_F^2 = \Tr \{ \Pb_k \Pb_k^H \} = 1$ we have
\begin{IEEEeqnarray}{rCl}
\label{eq_trsolution}
\Pb_k^{\timer} = \frac{ \Vtil_k^{(0)H} \Hbar_k}{\left\Vert \Vtil_k^{(0)H} \Hbar_k \right\Vert_F}.
\end{IEEEeqnarray}
Replacing (\ref{eq_trsolution}) in (\ref{eq_generalbd}) yields
\begin{IEEEeqnarray}{rCl}
\PP_k^{\timer} = \Vtil_k^{(0)} \frac{ \Vtil_k^{(0)H} \Hbar_k}{\left\Vert \Vtil_k^{(0)H} \Hbar_k \right\Vert_F}.
\end{IEEEeqnarray}

\section{EBD Precoder Solution}
\label{appB}
The EBD precoder, which operates as an equalizer at the transmitter, is found by solving
\begin{IEEEeqnarray}{rCl}
\label{eq_ebdproblemapp}
\min_{\Pb_k} \left\Vert \CC_k \Pb_k - \I_B \right\Vert_F^2,\quad \text{s.t. } \left\Vert \Vtil_k^{(0)}\Pb_k \right\Vert_F^2 = 1,
\end{IEEEeqnarray}
whose Lagrangian is
\begin{IEEEeqnarray}{rCl}
&& \lag_{\eq}\left(\Pb_k , \mu_k \right) \nonumber \\
&& = \Tr \left( \CC_k \Pb_k \Pb_k^H \CC_k^H \right) - \Tr \left( \CC_k \Pb_k \right) - \Tr \left( \Pb_k^H \CC_k^H \right) + B \nonumber \\
&& \quad + \mu_k \left[ \Tr \left( \Vtil_k^{(0)}\Pb_k \Pb_k^H \Vtil_k^{(0)H} \right)  - 1 \right]. \nonumber
\end{IEEEeqnarray}
Using the cyclic permutation invariance of the trace and $ \Vtil_k^{(0)H} \Vtil_k^{(0)} = \I_{B_c}$, the KKT condition for (\ref{eq_ebdproblemapp}) is
\begin{IEEEeqnarray}{rCl}
\label{eq_fonceq}
&& \frac{\partial \lag_{\eq}\left(\Pb_k , \mu_k \right)}{\partial \Pb_k} =  \CC_k^T \CC_k^* \Pb_k^* - \CC_k^T + \mu_k \Pb_k^* = \ze.
\end{IEEEeqnarray}
Using the complex-conjugate of (\ref{eq_fonceq}) we get $\Pb_k^{\eq} = \left( \CC_k^H \CC_k + \mu_k \I_{B_c} \right)^{-1} \CC_k^H$, which replacing into (\ref{eq_generalbd}) yields the EBD precoder
\begin{IEEEeqnarray}{rCl}
\PP_k^{\eq} & = &  \Vtil_k^{(0)} \left( \CC_k^H \CC_k + \mu_k \I_{B_c} \right)^{-1} \CC_k^H.
\end{IEEEeqnarray}
Note that, since $\CC_k^H \CC_k$ is Hermitian, the eigendecomposition $\CC_k^H \CC_k = \U_{C_k} \Lam_{C_k} \U_{C_k}^H$ is possible, where $\U_{C_k}$ is a unitary matrix and $\Lam_{C_k} = \diag\left( \lambda_{C_k,1}, \ldots, \lambda_{C_k,B_c}  \right)$ is the diagonal matrix with the (positive real) eigenvalues of $\CC_k^H \CC_k$. By enforcing the constraint $\Tr \left( \PP_k \PP_k^H \right) = 1$, we get
\begin{IEEEeqnarray}{rCl}
\label{eq_ebdlagrange}
&& \Tr \left( \PP_k^{\eq} \PP_k^{\eq \, H} \right) \nonumber \\
&& = \Tr \left( \left( \CC_k^H \CC_k + \mu_k \I_{B_c} \right)^{-1} \CC_k^H \CC_k \left( \CC_k^H \CC_k + \mu_k \I_{B_c} \right)^{-1} \right) \nonumber \\
&& = \Tr \left( \left( \Lam_{C_k} + \mu_k \I_{B_c} \right)^{-1} \Lam_{C_k} \left( \Lam_{C_k} + \mu_k \I_{B_c} \right)^{-1} \right) \nonumber \\
&& = \sum_{i=1}^{B_c} \frac{\lambda_{C_k,i}}{\left( \lambda_{C_k,i} + \mu_k \right)^2 } = 1.
\end{IEEEeqnarray}
Note that (\ref{eq_ebdlagrange}) has multiple solutions for $\mu_k$, but its left hand side is monotonically decreasing when $\mu_k \geq 0$. Thus, the unique solution for $\mu_k$ can be found by using a line search algorithm.

\section{JPBD Precoder Solution}
\label{appC}
We calculate the matrix $\Gu_k$ in (\ref{eq_jpsolution}) by maximizing the SNR at the receiver, which can be equivalently stated as
\begin{IEEEeqnarray}{rCl}
\label{eq_jpproblem}
\min_{\Gu_k} \left\Vert \Sig_k^+ \Gu_k^+ \right\Vert_F^2 \left\Vert \Gu_k \right\Vert_F^2,
\end{IEEEeqnarray}
with no constraints, since the precoder is already normalized. The first order necessary condition for this problem is
\begin{IEEEeqnarray}{rCl}
\label{eq_foncjp}
&& \frac{d}{d\Gu_k} \left( \left\Vert \Sig_k^+ \Gu_k^+ \right\Vert_F^2 \left\Vert \Gu_k \right\Vert_F^2 \right) \nonumber \\
&& = \left\Vert \Gu_k \right\Vert_F^2 \left ( -\Gu_k^{+T} \Sig_k^{+T}\Sig_k^+ \Gu_k^{+*} \Gu_k^{+T} + \Gu_k^{+T} \Gu_k^{+*} \Gu_k^{+T} \Sig_k^{+T}\Sig_k^+ \right. \nonumber \\
&& \quad \left. -  \Gu_k^{+T} \Gu_k^{+*} \Gu_k^{+T} \Sig_k^{+T}\Sig_k^+ \Gu_k^{+*} \Gu_k^{*} \right) + \left\Vert \Sig_k^+ \Gu_k^+ \right\Vert_F^2  \Gu_k^{*} \nonumber \\
&& = 0,
\end{IEEEeqnarray}
where we have used the complex matrix differentials defined in \cite{hjorungnes2007}. Applying complex-conjugate, using $\Gu_k^+ = \Gu_k^H \left(\Gu_k\Gu_k^H\right)^{-1}$, and rearranging (\ref{eq_foncjp}) gives
\begin{IEEEeqnarray}{rCl}
\left\Vert \Sig_k^+ \Gu_k^+ \right\Vert_F^2 \Gu_k \Gu_k^H = \left\Vert \Gu_k \right\Vert_F^2  \Gu_k^{+H} \Sig_k^{+H} \Sig_k^+ \Gu_k^+,
\end{IEEEeqnarray}
which is a nonlinear matrix equation with multiple stationary points for the objective function in (\ref{eq_jpproblem}). A general closed-form solution for this equation does not exist. Thus, for simplicity, assume $\Gu_k$ is rectangular diagonal with real positive entries $\gu_{k,i} = \left[ \Gu_k \right]_{ii},\, i=1,\ldots,B$. In such case, the objective function has the form
\begin{IEEEeqnarray}{rCl}
\label{eq_diagonalproblem}
\left\Vert \Sig_k^+ \Gu_k^+ \right\Vert_F^2 \left\Vert \Gu_k \right\Vert_F^2 = \left( \sum_{i=1}^B \frac{1}{\sigma_{k,i}^2 \gu_{k,i}^2} \right) \left( \sum_{i=1}^B \gu_{k,i}^2 \right) \geq \left( \sum_{i=1}^B \frac{1}{\sigma_{k,i}} \right)^2, \nonumber
\end{IEEEeqnarray}
where we applied the Cauchy-Schwarz inequality. Therefore, the objective function achieves its minimum when $\gu_{k,i} \propto 1/(\sigma_{k,i} \gu_{k,i})$ and we can define the closed-form solution
\begin{IEEEeqnarray}{rCl}
\gu_{k,i} = \sqrt{\frac{1}{\sigma_{k,i}}}.\nonumber
\end{IEEEeqnarray}

\section{Sum-Rate Maximization Solution}
\label{appD}
In this section, we show the solution to the power allocation problem for sum-rate maximization:
\begin{IEEEeqnarray}{rCl}
\label{eq_powerrateproblemapp}
\hspace*{-20pt}\max_{\rhob} \sum_{k'=1}^K \log_2 \left( 1 + \sinr_{k'} \right), \text{ s.t. }  \ \Vert \rhob \Vert_1 \leq P_{\max}, \ \rhob \geq 0.
\end{IEEEeqnarray}
The Lagrangian of (\ref{eq_powerrateproblemapp}) is 
\begin{IEEEeqnarray}{rCl}
&& \lag\left( \rhob , \lambda \right) = -\sum_{k'=1}^K \log_2 \left[ \frac{(\alpha_{\dd,k'}+\alpha_{\isi,k'})\rho_{k'}+\alpha_{\nn,k'}}{\alpha_{\isi,k'}\rho_{k'}+\alpha_{\nn,k'}} \right] \nonumber \\ 
&& \hspace*{50pt} + \lambda \left( \sum_{k'=1}^K \rho_{k'} - P_{\max} \right), \nonumber
\end{IEEEeqnarray}
where $\lambda \geq 0$ is a Lagrange multiplier. The KKT condition for this problem is
\begin{IEEEeqnarray}{rCl}
&& \frac{\partial \lag \left(\rhob , \lambda \right)}{\partial \rho_k} \nonumber \\
&& = -\frac{\alpha_{\dd,k} \alpha_{\nn,k}}{\ln (2) \left[ \left( \alpha_{\dd,k}+\alpha_{\isi,k} \right)\rho_k + \alpha_{\nn,k}\right] \left[ \alpha_{\isi,k} \rho_k + \alpha_{\nn,k} \right]} + \lambda \nonumber \\
&& = 0, \nonumber
\end{IEEEeqnarray}
which results in the following quadratic equation for $\rho_k$:
\begin{IEEEeqnarray}{rCl}
\rho_k & = & \frac{\pm \sqrt{\eta \alpha_{\dd,k}^2 \alpha_{\nn,k}^2 + \frac{4 \eta \alpha_{\dd,k} \alpha_{\isi,k} \alpha_{\nn,k}}{\lambda \ln(2)}(\alpha_{\dd,k}+\alpha_{\isi,k})} } {2 \alpha_{\isi,k} (\alpha_{\dd,k}+\alpha_{\isi,k})} \nonumber \\
&& - \frac{\eta \alpha_{\nn,k} (\alpha_{\dd,k}+2\alpha_{\isi,k})}{2 \alpha_{\isi,k} (\alpha_{\dd,k}+\alpha_{\isi,k})},
\end{IEEEeqnarray}
Note that the above equation has two solutions for every $k$, so we select the positive sign in the first factor, which gives a positive power. Thus, enforcing both $\sum_k \rho_k \leq P_{\max}$ and $\rhob \geq 0$ gives
\begin{IEEEeqnarray}{rCl}
\label{eq_poweralloclagrange}
\hspace*{-20pt} \sum_{k=1}^K \frac{\sqrt{\eta^2 \alpha_{\dd,k}^2 \alpha_{\nn,k}^2 + \frac{4 \eta \alpha_{\dd,k} \alpha_{\isi,k} \alpha_{\nn,k}}{\lambda \ln(2)}(\alpha_{\dd,k}+\alpha_{\isi,k})}}{2 \alpha_{\isi,k} (\alpha_{\dd,k}+\alpha_{\isi,k})} && \nonumber \\
\hspace*{-20pt} - \sum_{k=1}^K \frac{\eta \alpha_{\nn,k} (\alpha_{\dd,k}+2\alpha_{\isi,k})}{2 \alpha_{\isi,k} (\alpha_{\dd,k}+\alpha_{\isi,k})} & = & P_{\max}.
\end{IEEEeqnarray}
where $0 \leq \lambda \leq \alpha_{\dd,k}/[\alpha_{\nn,k} \ln(2)]$, $\forall k$, must hold so the power allocated to each user is positive. Note that the right hand side of (\ref{eq_poweralloclagrange}) is monotonically decreasing on $\lambda$, and hence a unique solution to (\ref{eq_poweralloclagrange}) can be found through a line search over the interval
\begin{IEEEeqnarray}{rCl}
0 \leq \lambda \leq \min_k \frac{\alpha_{\dd,k}}{ \alpha_{\nn,k} \ln(2)}.
\end{IEEEeqnarray}